# THERMAL CONDUCTIVITY OF OXIDE SCALE AND ITS COMPONENTS IN THE RANGE FROM 0 ºC TO 1300 ºC: GENERALIZED ESTIMATES WITH ACCOUNT FOR MOVABILITY OF PHASE TRANSITIONS

*Emmanuil Beygelzimer[1], Yan Beygelzimer[2]*
[1]*OMD-Engineering LLC, Dnipro, Ukraine*
[2]*Donetsk Institute for Physics and Engineering named after A.A. Galkin, National Academy of Sciences of Ukraine, Kyiv, Ukraine*

*Corresponding author: emmanuilomd@gmail.com, tel.: +380 (50) 368-63-42, 49000, Volodymyr Monomakh Street, 6 of. 303, Dnipro, Ukraine.

**ABSTRACT**

The data of different authors on the thermal conductivity of wüstite $Fe_{1-x}O$, magnetite $Fe_3O_4$, hematite $Fe_2O_3$ and pure iron are systematized. The generalized values are described by piecewise smooth functions containing as varying parameters the temperatures of magnetic and polymorphic (for iron) transformations as well as the thermodynamic stability boundary (for wüstite). At polymorphic transformation a finite break of the thermal conductivity function is envisaged, at other critical points only a break of its temperature derivative is acceptable. The proposed formulas are presented in two forms: the general form that allow varying the values of critical temperatures and the particular form corresponding to their basic values: the boundary of thermodynamic stability of wüstite - 570 ºC, the Curie points of magnetite - 575 ºC, of hematite - 677 ºC, of iron – 770 ºC and polymorphic transformation temperature of iron - 912ºC. To calculate the thermal conductivity of oxide scale as a whole, it is proposed to take into account separately metallic iron and composite matrix of iron oxides. Model computations using the proposed formulas shows that the true thermal conductivity of oxide scale (without pores), depending on the temperature may be from 3 to 6 $W·m^{-1}·K^{-1}$ in the absence of metallic iron and up to 15 $W·m^{-1}·K^{-1}$ if free iron is released in the eutectoid decay of wüstite. The effective thermal conductivity of oxide scale, taking into account its real porosity, may be lower by 15-35%. The obtained dependencies are recommended for use in mathematical simulation of production and processing of steel products in the presence of oxide scale on their surface.

**INTRODUCTION**

This report completes a series of publications of the authors on the mathematical description of the thermophysical properties of wüstite $Fe_{1-x}O$ ($x \approx 0.05...0.15$), magnetite $Fe_3O_4$, hematite $Fe_2O_3$, and metallic iron Fe, as well as oxide scale in general in the temperature range from 0 °C to 1300 °C. The thermal expansion coefficient [1], density [2] and heat capacity [3] have been described earlier. In this paper, we look at the thermal conductivity coefficient, which, as usual, is understood as the coefficient of proportionality between the vector of heat flux density and the temperature gradient.

When approximating the known data on thermophysical properties, the authors set themselves two main tasks (see [1]): 1) to consider the movability of phase transitions in order to use it for adaptation of mathematical models, and 2) to generalize the results of various studies in order to recommend the most characteristic values of the properties at different temperatures for a given substance. The first task was solved by including the temperatures of phase transitions in the approximating functions as *varying parameters*. The second problem (generalization) was solved by choosing the *type* of approximating function and the coordinates of the *reference points* through which the graph of this function should pass. In this respect, an essential feature of the studies of thermal conductivity presented in this article is related to the fact that the empirical data on this characteristic have the largest scatter among all thermophysical properties. For example, for any of the considered iron oxides, the data of different authors on the thermal conductivity may differ by 2-4 times. In this regard, the choice of coordinates of reference points for the thermal conductivity coefficient is to a large extent conditional, because it is possible that new empirical data that will be obtained in the future, or the thermal conductivity values in some specific conditions may not correspond to the choice made by the authors. However, the authors hope that for each of the considered substances the chosen type of approximating functions correctly reflects the nature of the dependence of thermal conductivity on temperature, and the accumulation of new empirical data may require only a correction of the coordinates of reference points, but not a revision of the approximating functions.

**METHODS**

Approximation of data on the thermal conductivity of the structural components of oxide scale was carried out according to the general methods described in detail in the first part [1]. In accordance with these methods, the critical temperatures are included in the approximating functions in the form of formal parameters. Such critical temperatures are taken as: Curie points (for $Fe_3O_4$, $Fe_2O_3$, Fe), polymorphic transformation temperature (for Fe), thermodynamic stability boundary (Chaudron point for $Fe_{1-x}O$). The real ranges of movability of these critical temperatures and their basic values, adopted for a particular form of approximating formulas, are given in [1]. As a key assumption, it is considered that if the critical temperature changes within the real range of its movability, the values of thermal conductivity at this critical point and at two distant reference points on different sides of it remain unchanged. In other words, it is assumed that as the critical point changes, the temperature dependence graph of the thermal conductivity shifts parallel to the temperature axis, remaining fixed at the extreme reference points. Accordingly, when the critical temperature varies, the other parameters of the approximating functions change automatically.

Conjugation of approximating formulas between intervals separated by critical points was performed taking into account the accepted "behavior model" of thermal conductivity in critical states [1]: with a break of the function during the polymorphic transformation and with a break of the first derivative function (but without breaking the thermal conductivity coefficient itself) at the Curie points and at the boundary of thermodynamic stability.

The type of approximating formulas was chosen according to the value of specific thermal resistance $k$ [$m·K·W^{-1}$], the inverse of the thermal conductivity coefficient $\lambda$ [$W·m^{-1}·K^{-1}$] (it turned out to be more convenient, the idea was borrowed from [5].):

$$k = \frac{1}{\lambda} \qquad (1)$$

Therefore, the data on the thermal conductivity coefficient known from the technical literature was first converted into the values of specific thermal resistance and only then approximated. Accordingly, the ordinates of all reference points for the approximating function were expressed in terms of the thermal resistance. The reverse recalculation into the thermal conductivity coefficient was performed at the very end after obtaining the approximating function for the resistivity.

Taking into account the large scatter of currently available empirical data (see Introduction), the coordinates of the reference points chosen by the authors should be considered as "*basic*". In the case of new data, the values of these coordinates can be adjusted without changing the general formulas proposed.

When processing the information on iron oxides we focused on the data obtained on polycrystalline samples, and with minimal porosity (or recalculated to zero porosity, as, for example, in [4]). Data on single crystals were used as additional information.

## RESULTS AND DISCUSSION

**Magnetite $Fe_3O_4$**

Data on the thermal conductivity coefficient of $Fe_3O_4$ available in the technical literature are shown in **Fig. 1**. The approximation was performed separately in two temperature intervals: below and above the Curie point. The values of the thermal conductivity of magnetite from [5] were not taken into account, since they are drastically different from other data. The resulting generalized formulas are summarized in **Table 1** and contain the Curie point ($T_1$) as a formal parameter.
At the basic value of the Curie point of magnetite $T_1$ = 848 K (575 °C), formulas from Table 1 are reduced to the particular form:
– in the range of 273 K ≤ $T$ ≤ 848 K

$$\lambda_{Fe3O4} = (0{,}10136 + 2{,}9321 \cdot 10^{-4} \cdot T - 0{,}10165 \cdot T^{-2})^{-1} \quad (2)$$

– in the range of 848 K < $T$ ≤ 1573 K

$$\lambda_{Fe3O4} = 2{,}857 \quad (3)$$

where $\lambda_{Fe3O4}$ [W·m$^{-1}$·K$^{-1}$] is the thermal conductivity coefficient of magnetite; $T$ [K] is the design temperature.
The graph built by formulas (2)-(3) is shown in Fig. 1. Another graph (denoted "Approx 2") in the same figure built by the general formulas from Table 1 with the same basic value of the Curie point of magnetite, but with other values of reference points coordinates, namely: $T_0$ = 200 K; $k_0$ = 0.14 m·K·W$^{-1}$, $k_1$ = 0.28 m·K·W$^{-1}$, $T_2$ = 1600 K; $k_2$ = 0.24 m·K·W$^{-1}$ (if we take into account the data of measurements in work [5] and the tendency to a slight rise in the thermal conductivity coefficient with increasing temperature above the Curie point, noted in [6; 10]). However, it should be emphasized that this graph is given only to demonstrate the possibility of the proposed approach; to date, in the absence of other data, for generalized estimates of the thermal conductivity of magnetite the authors recommends using the basic values of the coordinates of the reference points given in Table 1.
According to the formulas of Table 1, the shift of the Curie point of magnetite within the real limits of its movability can change its thermal conductivity by no more than 6-7%. For example, at 500 °C the calculated value of $\lambda_{Fe3O4}$ is 2.99 W·m$^{-1}$K$^{-1}$ if the magnetic transition occurs at $T_1$ = 823 K (550 °C), and 3.17 W·m$^{-1}$K$^{-1}$ if the Curie point is $T_1$ = 900 K (627 °C).

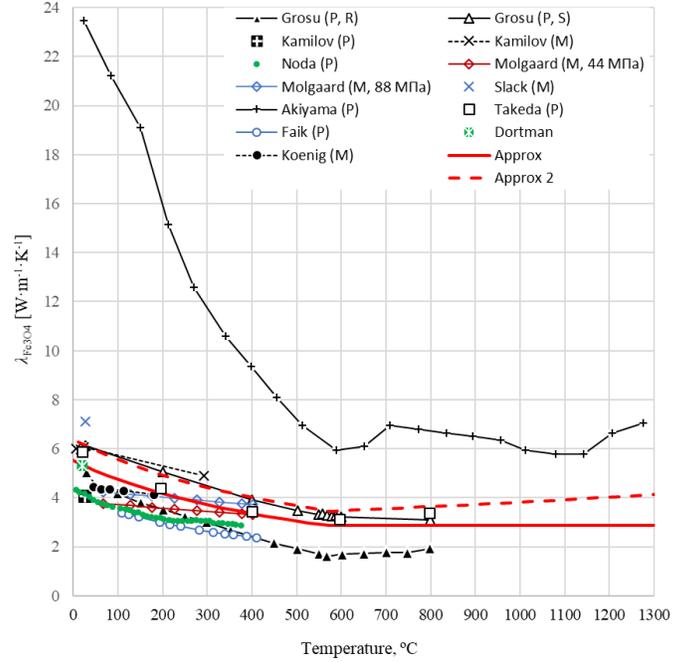

**Fig. 1.** Thermal conductivity coefficient of magnetite ($Fe_3O_4$) according to empirical data of Grosu [6], Kamilov [7], Noda [4], Molgaard (separately at pressures 44 and 88 MPa) [8], Slack [9], Akiyama [5], Takeda [10], Faik [11], Dortman [12], Koenig [13, p. 154, array 4]. The conventional symbols in the legend of the graph: M - monocrystal, P - polycrystal, R - raw ore, S - sintered sample. The experimental points are conventionally connected by straight lines. "Approx" denotes the graph built according to the formulas (2)-(3) (with the basic value of the Curie point of magnetite $T_1$ = 848 K (575 °C) and coordinates of the reference points from Table 1), "Approx 2" is the graph built according to the formulas from Table 1 with the basic value of the Curie point but with alternative coordinates of the reference points – see text.

**Table 1.** Formulas for the thermal conductivity coefficient of $Fe_3O_4$ in two temperature intervals ($T_1$ [K] is the Curie point of magnetite)

| Temperature interval [K] | $273 \leq T \leq T_1$ | | |
|---|---|---|---|
| Approximating function [W·m$^{-1}$K$^{-1}$] | $\lambda_{Fe3O4} = (a_0 + a_1 T^n + a_2 T^m)^{-1}$ | | |
| Constants | $n = 1.0$ | | $m = -2.0$ |
| Basic coordinates of the reference points | $T_0 = 200$ K | $k_0 = 0.16$ m·K·W$^{-1}$ | $k_1 = 0.35$ m·K·W$^{-1}$ |
| Coefficients to be calculated | $a_0 = \dfrac{k_1 T_0^m + k_0 T_1^n - k_0 T_1^m - k_1 T_0^n}{T_0^n T_1^m - T_1^n T_0^m - T_1^m + T_0^m + T_1^n - T_0^n}$ | | |
| | $a_1 = \dfrac{k_0 T_1^m - k_1 T_0^m - a_0 (T_1^m - T_0^m)}{T_0^n T_1^m - T_1^n T_0^m}$ | | |
| | $a_2 = \dfrac{k_0 T_1^n - k_1 T_0^n - a_0 (T_1^n - T_0^n)}{T_0^m T_1^n - T_1^m T_0^n}$ | | |
| Temperature interval [K] | $T_1 < T \leq 1573$ | | |
| Approximating function [W·m$^{-1}$K$^{-1}$] | $\lambda_{Fe3O4} = (b_0 + b_1 (T - T_1))^{-1}$ | | |
| Basic coordinates of the reference points | $T_2 = 1600$ K | $k_2 = 0.35$ m·K·W$^{-1}$ | |
| Coefficients to be calculated | $b_0 = k_1$ | | |
| | $b_1 = \dfrac{k_2 - k_1}{T_2 - T_1}$ | | |

## Wüstite $Fe_{1-x}O$

Data on the thermal conductivity coefficient of wüstite available in the technical literature are shown in **Fig. 2**. Taking into account the results of [5], the approximation was performed separately in two temperature intervals: before and after the Chaudron point (the authors of [5] correlated the kink in the temperature dependence of the thermal conductivity of wustite not with the Chaudron point, but with its Tammann temperature (825 K/552 °C), equal to half of the melting point in Kelvin degrees). **Table 2** shows the formulas in general form with the Chaudron point $T_1$ as a formal parameter. At the basic value of $T_1 = 843$ K (570 °C) these formulas take the particular form (the corresponding graph is shown in Fig. 2):

– in the range of $273\ K \le T \le 843\ K$

$$\lambda_{FeO} = (2.7054 \cdot 10^{-2} + 9.4008 \cdot 10^{-3} \cdot T^{0.5} - 3.6455 \cdot T^{-2})^{-1} \quad (4)$$

– in the range of $843\ K < T \le 1573\ K$

$$\lambda_{FeO} = \left(0.3 - 7.9260 \cdot 10^{-5}(T - 843)\right)^{-1} \quad (5)$$

where $\lambda_{FeO}$ [W·m$^{-1}$·K$^{-1}$] is the thermal conductivity coefficient of wüstite; $T$ [K] is the design temperature.

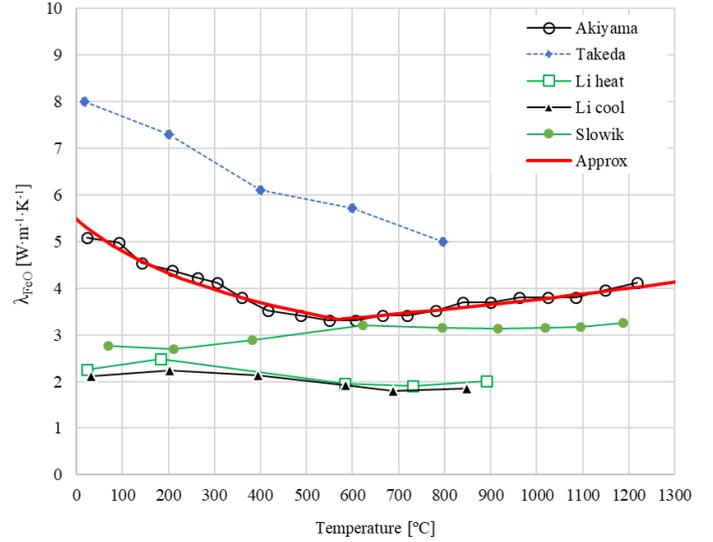

**Fig. 2.** Thermal conductivity coefficient of wüstite according to Akiyama [5], Takeda [10], Li (separately for heating and cooling) [14, p. 2104], Slowik [15]. "Approx" - graph built by approximating formulas (4)-(5).

**Table 2.** Formulas for calculating the thermal conductivity coefficient of wüstite in two temperature intervals ($T_1$ [K] is the Chaudron point)

| Temperature interval [K] | $273 \le T \le T_1$ | | |
|---|---|---|---|
| Approximating function [W·m$^{-1}$K$^{-1}$] | $\lambda_{FeO} = (a_0 + a_1 T^n + a_2 T^m)^{-1}$ | | |
| Constants | $n = 0.5$ | | $m = -2.0$ |
| Basic coordinates of the reference points | $T_0 = 200$ K | $k_0 = 0.16$ m·K·W$^{-1}$ | $k_1 = 0.3$ m·K·W$^{-1}$ |
| Coefficients to be calculated | $a_0 = \dfrac{k_1 T_0^m + k_0 T_1^n - k_0 T_1^m - k_1 T_0^n}{T_0^n T_1^m - T_1^n T_0^m - T_1^m + T_0^m + T_1^n - T_0^n}$ | | |
| | $a_1 = \dfrac{k_0 T_1^m - k_1 T_0^m - a_0(T_1^m - T_0^m)}{T_0^n T_1^m - T_1^n T_0^m}$ | | |
| | $a_2 = \dfrac{k_0 T_1^n - k_1 T_0^n - a_0(T_1^n - T_0^n)}{T_0^m T_1^n - T_1^m T_0^n}$ | | |
| Temperature interval [K] | $T_1 < T \le 1573$ | | |
| Approximating function [W·m$^{-1}$K$^{-1}$] | $\lambda_{FeO} = \left(b_0 + b_1(T - T_1)\right)^{-1}$ | | |
| Basic coordinates of the reference points | $T_2 = 1600$ K | | $k_2 = 0.24$ m·K·W$^{-1}$ |
| Coefficients to be calculated | $b_0 = k_1$ | | |
| | $b_1 = \dfrac{k_2 - k_1}{T_2 - T_1}$ | | |

## Hematite $Fe_2O_3$

Known from the technical literature data on the thermal conductivity of hematite are shown in **Fig. 3**. The approximating formulas are summarized in **Table 3** and contain the Curie point of hematite as a formal parameter $T_1$. At the basic value of $T_1 = 950$ K (677 °C) for hematite, these formulas are reduced to the particular form (the corresponding graph is shown in Fig. 3):

– in the range of $273\ K \le T \le 950\ K$

$$\lambda_{Fe2O3} = \left(0.25 - 2.6667 \cdot 10^{-4}(950 - T)\right)^{-1} \quad (6)$$

– in the range of $950\ K < T \le 1573\ K$

$$\lambda_{Fe2O3} = \left(0.25 + 6.1538 \cdot 10^{-5}(T - 950)\right)^{-1} \quad (7)$$

where $\lambda_{Fe2O3}$ [W·m$^{-1}$·K$^{-1}$] is the thermal conductivity coefficient of hematite; $T$ [K] is the design temperature.

According to the formulas in Table 3, hematite has about the same sensitivity of the thermal conductivity to Curie point movability as that of magnetite, namely no more than 6%. For example, at Curie point $T_1 = 943$ K (670 ºC) the calculated value of $\lambda_{Fe2O3}$ at this temperature is 4.00 W·m$^{-1}$·K$^{-1}$; if the magnetic transition occurs at $T_1 = 998$ K (725 ºC), then the calculated value of $\lambda_{Fe2O3}$ at the same temperature of 670 ºC is 4.23 W·m$^{-1}$·K$^{-1}$.

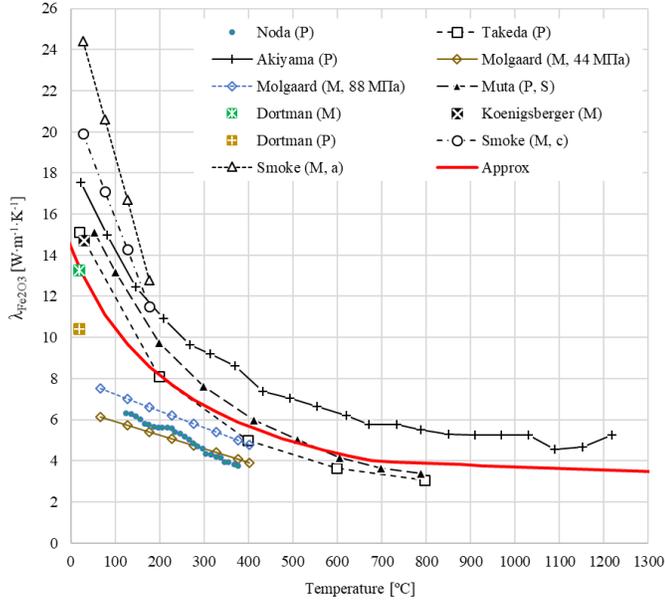

**Fig. 3**. Thermal conductivity coefficient of hematite according to empirical data from Noda [4], Takeda [10], Akiyama [5], Molgaard (separately at pressures 44 and 88 MPa) [8], Muta [6], Dortman [12], Koenigsberger (cited from [4]) and Smoke [16, p. 64]. Notation: M - monocrystal (a, c - crystal axes), P - polycrystal, S - sintered. The experimental points are conventionally connected by straight lines. The graph built by formulas (6)-(7) is shown (denoted as "Approx")

**Iron Fe**

Experimental data of different authors on the thermal conductivity coefficient of iron are systematized in **Fig. 4**. Approximating formulas in general form are given in **Table 4** and contain the following defining parameters: $T$ [K] is design temperature; $T_1$ [K] is the Curie point of iron; $T_2$ [K] is the temperature of polymorphic transformation ($\gamma \to \alpha$ transition at cooling).

Calculations by formulas of Table 4 show that shifting the Curie point within its real movability range can change the thermal conductivity coefficient of iron by no more than 4%. For example, the calculated value of $\lambda_{Fe}$ at 750 ºC is 29.6 W·m$^{-1}$·K$^{-1}$ if the Curie point is $T_1 = 1032$ K (759 ºC) and 30.7 W·m$^{-1}$·K$^{-1}$ if $T_1 = 1046$ K (773 ºC).

The sensitivity of the thermal conductivity coefficient of iron to the variations of the polymorphic transformation temperature $T_2$ is quite obviously determined by the jump of $\lambda_{Fe}$ at point $T_2$ from 30.0 W·m$^{-1}$·K$^{-1}$ in $\alpha$-phase (below $T_2$) to 27.7 W·m$^{-1}$·K$^{-1}$ in $\gamma$-phase (above $T_2$).

At the basic values of iron Curie point $T_1 = 1043$ K (770ºC) and polymorphic transformation $T_2 = 1185$ K (912ºC) formulas from Table 4 are reduced to the particular form (corresponding graph shown in Figure 4):

– in the range of 273 K $\leq T \leq$ 1043 K

$$\lambda_{Fe} = [7.7 \cdot 10^{-3} + 9.2122 \cdot 10^{-6} \cdot T^{1.11} + 6.4624 \cdot 10^{-3} e^{-0.014 \cdot (1043-T)}]^{-1} \quad (8)$$

– in the range of 1043 K $< T \leq$ 1185 K

$$\lambda_{Fe} = [3.3295 \cdot 10^{-2} + 1.5051 \cdot 10^{-3} e^{-0.04 \cdot (T-1043)}]^{-1} \quad (9)$$

– in the range of 1185 K $< T \leq$ 1573 K:

$$\lambda_{Fe} = [2.7804 \cdot 10^{-2} + 1.6359 \cdot 10^{10} \cdot T^{-4}]^{-1} \quad (10)$$

where $\lambda_{Fe}$ [W·m$^{-1}$·K$^{-1}$] is the thermal conductivity coefficient of iron; $T$ [K] is the design temperature.

**Table 3.** Formulas for calculating the thermal conductivity coefficient of hematite in two temperature intervals ($T_1$ [K] is the Curie point)

| Temperature interval [K] | $273 \leq T \leq T_1$ | | |
|---|---|---|---|
| Approximating function [W·m$^{-1}$K$^{-1}$] | $\lambda_{Fe2O3} = (a_0 + a_1(T_1 - T))^{-1}$ | | |
| Basic coordinates of the reference points | $T_0 = 200$ K | $k_0 = 0.05$ m·K·W$^{-1}$ | $k_1 = 0.25$ m·K·W$^{-1}$ |
| Coefficients to be calculated | $a_0 = k_1$ | | |
| | $a_1 = \dfrac{k_0 - k_1}{T_1 - T_0}$ | | |
| Temperature interval [K] | $T_1 < T \leq 1573$ K | | |
| Approximating function [W·m$^{-1}$K$^{-1}$] | $\lambda_{Fe2O3} = (b_0 + b_1(T - T_1))^{-1}$ | | |
| Basic coordinates of the reference points | $T_2 = 1600$ K | | $k_2 = 0.29$ m·K·W$^{-1}$ |
| Coefficients to be calculated | $b_0 = k_1$ | | |
| | $b_1 = \dfrac{k_2 - k_1}{T_2 - T_1}$ | | |

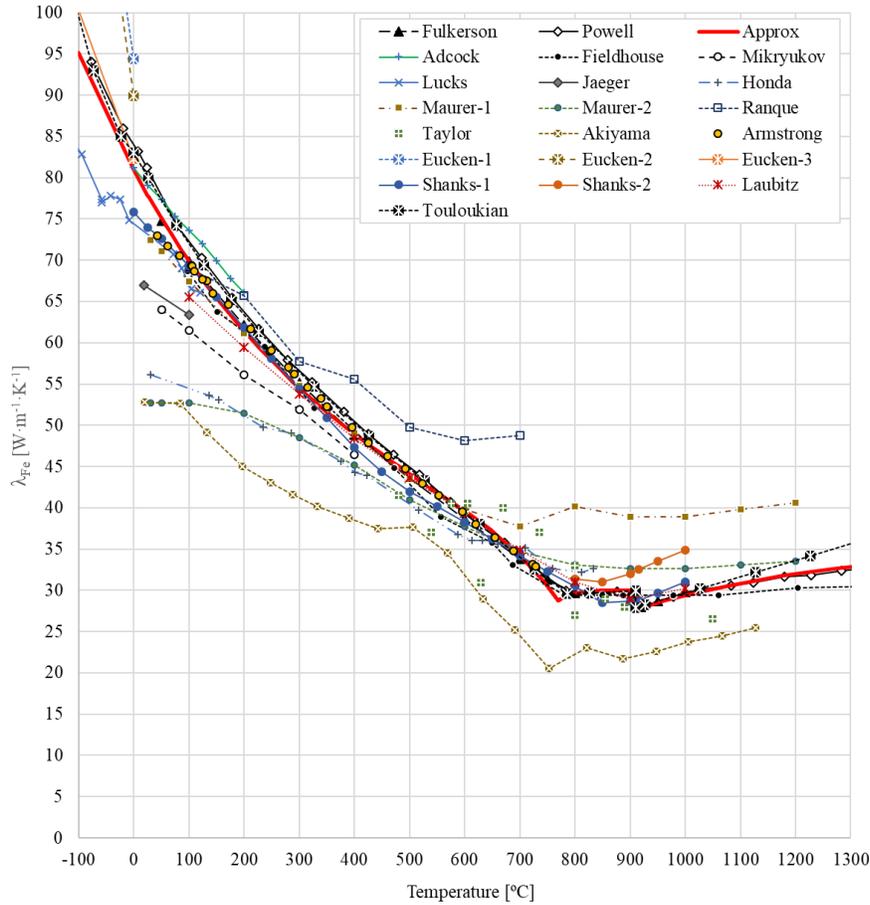

**Fig. 4.** Thermal conductivity coefficient of pure iron according to Fulkerson [17], Powell [18, p. 35], Adcock [19, p. 156, set 3], Fieldhouse [19, p. 156, set 6], Mikrukov [20], Lucks [19, p. 156, set 12], Jaeger [19, p. 156, set 15], Honda [19, p. 156, set 24], Maurer [19, c. 156, sets 34 and 35], Ranque [19, c. 156, set 47], Taylor [19, c. 156, set 62], Akiyama [5], Armstrong [21], Eucken [19, c. 156, sets 7, 8 and 9], Shanks [22], Laubitz [23] and the generalized recommendations of Touloukian [16, p. 15]. "Approx" is the graph calculated by (8)-(10).

**Oxide scale as a whole**

**Fig. 5** compares temperature dependences of thermal conductivity coefficient of different structural components of oxide scale calculated according to the above formulas for basic values of critical temperatures. According to these generalized data the thermal conductivity of iron oxides in the temperature range from 0 to 1300 °C is approximately 10 times lower than that of iron itself. At temperatures from 600 to 1300 °C thermal conductivities of all iron oxides are approximately equal, and in the range from 0 to 600 °C thermal conductivity of hematite is approximately 1.5-3 times higher than that of wüstite and magnetite. For oxide scale in general, two aspects should be considered: 1) the effect of each solid component of scale on its overall thermal conductivity and 2) the porosity of oxide scale.

In order to assess the effect of the individual components of the oxide scale on its overall thermal conductivity, we first consider a composite material consisting of only two components. The component with the higher thermal conductivity will be assigned an ordinal number "1". In other words, we will assume that:

$$\lambda_1 \geq \lambda_2 \tag{11}$$

where $\lambda_1$ and $\lambda_2$ are thermal conductivity coefficients of the components.
Let us denote the volume fractions of the components by symbol $\varphi$ with the corresponding index, whereby:

$$\varphi_1 + \varphi_2 = 1 \tag{12}$$

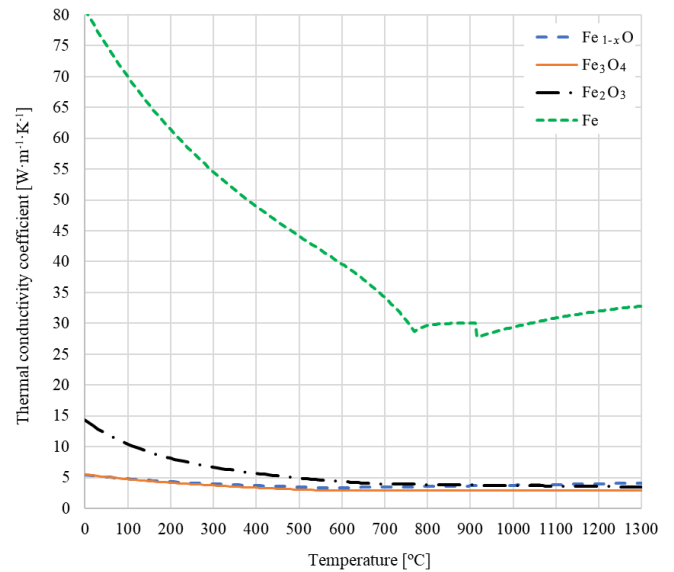

**Fig. 5.** Comparison of thermal conductivity coefficient of wüstite, magnetite, hematite and iron depending on temperature. Calculations were made by formulas (4)-(5), (2)-(3), (6)-(7) and (8)-(10), respectively.

**Table 4.** Formulas for the thermal conductivity coefficient of iron ($T_1$ [K] is the Curie point, $T_2$ [K] is the temperature of polymorphic transformation)

| Temperature interval [K] | $273 \leq T \leq T_1$ | | |
|---|---|---|---|
| Approximating function [W·m⁻¹K⁻¹] | $\lambda_{\text{Fe}} = \left(a_0 + a_1 T^n + a_3 e^{-a_4(T_1-T)}\right)^{-1}$ | | |
| Constants | $n = 1.11$ | $a_0 = 0.0077$ | $a_4 = 0.014$ |
| Basic coordinates of the reference points | $T_0 = 200$ K | $k_0 = 0.011$ m·K·W⁻¹ | $k_1 = 0.0348$ m·K·W⁻¹ |
| Coefficients to be calculated | $a_1 = \dfrac{W_1(k_1 - a_0) + a_0 - k_0}{W_1 T_1^n - T_0^n}$ | | |
| | $a_3 = k_1 - a_0 - a_1 T_1^n$ | | |
| Auxiliary parameter | $W_1 = e^{-a_4(T_1 - T_0)}$ | | |
| Temperature interval [K] | $T_1 < T \leq T_2$ | | |
| Approximating function [W·m⁻¹K⁻¹] | $\lambda_{\text{Fe}} = \left(b_0 + b_3 e^{-b_4(T-T_1)}\right)^{-1}$ | | |
| Constants | $b_4 = 0.04$ | | |
| Basic coordinates of the reference points | $k_{2(2)} = 0.0333$ m·K·W⁻¹ | | |
| Coefficients to be calculated | $b_3 = \dfrac{k_1 - k_{2(2)}}{1 - W_2}$ | | |
| | $b_0 = k_1 - b_3$ | | |
| Auxiliary parameter | $W_2 = e^{-b_4(T_2 - T_1)}$ | | |
| Temperature interval [K] | $T_2 < T \leq 1573$ | | |
| Approximating function [W·m⁻¹K⁻¹] | $\lambda_{\text{Fe}} = (d_0 + d_1 T^p)^{-1}$ | | |
| Constants | $p = -4$ | | |
| Basic coordinates of the reference points | $T_3 = 1600$ K | $k_{2(3)} = 0.0361$ m·K·W⁻¹ | $k_3 = 0.0303$ m·K·W⁻¹ |
| Coefficients to be calculated | $d_1 = \dfrac{k_3 - k_{2(3)}}{T_3^p - T_2^p}$ | | |
| | $d_0 = k_{2(3)} - d_1 T_2^p$ | | |

It is known that the total thermal conductivity of a composite lies between two limit values which correspond to the cases of parallel and sequential arrangement of components relative to the direction of the heat flow [24]. In the "parallel connection" model, i.e., when the components are arranged in such a way that the boundary surface between them is parallel to the heat flow, the composite thermal conductivity $\lambda$ is determined as the arithmetic mean value (this is the upper estimate):

$$\lambda = \lambda_1 \varphi_1 + \lambda_2 \varphi_2 \tag{13}$$

In the "sequential connection" model, i.e., when the components are arranged one after the other across the heat flow, the thermal resistances are averaged (this is the lower estimate):

$$\frac{1}{\lambda} = \frac{\varphi_1}{\lambda_1} + \frac{\varphi_2}{\lambda_2} \tag{14}$$

For composite materials in which one of the components is present in the form of chaotically arranged closed inclusions in a matrix of the second component, in practice the formula of Odelevski [25; 26, p. 24] is applied, which gives a result intermediate between models (13) and (14):

$$\frac{\lambda}{\lambda_1} = 1 - \frac{\varphi_2}{\dfrac{\lambda_1}{\lambda_1 - \lambda_2} - \dfrac{\varphi_1}{3}} \tag{15}$$

For further analysis it is convenient to enter relative parameters:
− the relative heat transfer coefficient of the second component:

$$Y_2 = \frac{\lambda_2}{\lambda_1} \leq 1,0 \tag{16}$$

− relative thermal conductivity coefficient of the composite material

$$Y = \frac{\lambda}{\lambda_1} \leq 1,0 \tag{17}$$

In these relative parameters, the above calculated models of the total thermal conductivity of the composite are written as follows:
− parallel connection model (upper estimate, analogous to (13)):

$$Y = \varphi_1 + Y_2(1 - \varphi_1) \tag{18}$$

− sequential connection model (lower estimate, analogous to (14)):

$$Y = \left[\varphi_1 + \frac{1-\varphi_1}{Y_2}\right]^{-1} \qquad (19)$$

– Odelevski model (analogous to (15)):

$$Y = 1 - \frac{1-\varphi_1}{\frac{1}{1-Y_2} - \frac{\varphi_1}{3}} \qquad (20)$$

**Fig. 6** shows the graphs built by formulas (18)-(20). They give a visual assessment of the influence of the relative thermal conductivity and the volume fraction of the components at different calculated models of the total thermal conductivity of the composite. It can be seen that if the thermal conductivity coefficients of the individual components are close to each other ($Y_2 = 0.9$ corresponds to this case in Fig. 6), then all models give approximately the same composite thermal conductivity result. If the thermal conductivities of the components differ by an order of magnitude (the case $Y_2 = 0.1$ in Fig. 6), then the range of results between the limiting models becomes very large, and the calculated estimate of the total thermal conductivity of the composite significantly depends on the selected model. It is indicative that the sequential model turns out to be insensitive to a small volume fraction of the first component, and within its content up to 20% ($\varphi_1 < 0.2$) is determined by the least thermally conductive component.

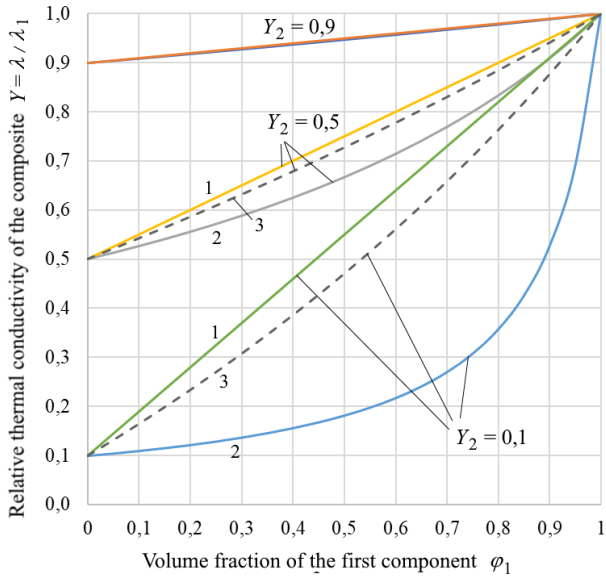

**Fig. 6.** Calculated models of the relative thermal conductivity of the composite as a function of the volume fraction of the component with the highest thermal conductivity: 1 - parallel connection model; 2 - sequential connection model; 3 - Odelevski model. $Y_2$ is the relative thermal conductivity of the second component (see (16)). For $Y_2 = 0.9$ the model numbers are not shown, since the graphs are practically the same.

Applied to our case of surface oxide scale, two of the considered models may be relevant: sequential bonding and Odelevski one. Indeed, the oxides are characterized by a layered arrangement on the surface (with increasing oxygen content as the distance from the metal increases, for example, an inner layer of $Fe_{1-x}O$, above it - $Fe_3O_4$ and an outer layer of $Fe_2O_3$). Since the heat flux is usually directed perpendicular to the surface, this arrangement of oxides corresponds to a sequential pattern of overall thermal conductivity of the oxide scale. The situation is quite different when metallic iron is present in the oxide scale, for example, due to eutectoid decay of wüstite. In this case, the iron particles can be regarded as chaotically arranged closed inclusions, which corresponds to the Odelevski model. On this basis, to calculate the thermal conductivity of oxide scale in the general case, we propose a "combined" model, in which the influence of metallic iron is accounted for by the Odelevski formula, and oxides - by the model of sequential connection. In the notations we have adopted, this combined model is written in the form of the following chain of formulas:

1) Oxide volume fractions excluding metallic iron

$$\omega_{FeO} = \frac{\psi_{FeO}}{1-\psi_{Fe}}; \; \omega_{Fe3O4} = \frac{\psi_{Fe3O4}}{1-\psi_{Fe}}; \; \omega_{Fe2O3} = \frac{\psi_{Fe2O3}}{1-\psi_{Fe}};$$
$$\omega_{xO} = \frac{\psi_{xO}}{1-\psi_{Fe}} \qquad (21)$$

where the symbol $\omega$ with the index denotes the volume fraction of the corresponding oxide in the solid material of oxide scale without metallic iron (the index "$x$O" means the oxide of the alloying element); the symbol $\psi$ with the index denotes the volume fraction of the corresponding component in the solid material of oxide scale, including metallic iron. The indicated volume fractions meet the conditions of rationing:

$$\psi_{FeO} + \psi_{Fe3O4} + \psi_{Fe2O3} + \psi_{xO} + \psi_{Fe} = 1 \qquad (22)$$

and

$$\omega_{FeO} + \omega_{Fe3O4} + \omega_{Fe2O3} + \omega_{xO} = 1 \qquad (23)$$

2) Cumulative thermal conductivity coefficient of the oxide matrix (according to the model of sequential connection):

$$\lambda_{ox} = \left[\frac{\omega_{FeO}}{\lambda_{FeO}} + \frac{\omega_{Fe3O4}}{\lambda_{Fe3O4}} + \frac{\omega_{Fe2O3}}{\lambda_{Fe2O3}} + \frac{\omega_{xO}}{\lambda_{xO}}\right]^{-1} \qquad (24)$$

where in the right part, the symbol $\lambda$ with an index indicates the thermal conductivity coefficient of the corresponding oxide (for example, for iron oxides, according to the formulas given earlier in this article);

3) Total heat transfer coefficient of oxide scale without pores (according to Odelevski model):

$$\lambda_{sc} = \lambda_{Fe}\left[1 - \frac{1-\psi_{Fe}}{\frac{\lambda_{Fe}}{\lambda_{Fe}-\lambda_{ox}} - \frac{\psi_{Fe}}{3}}\right] \qquad (25)$$

where $\lambda_{Fe}$ is the thermal conductivity coefficient of iron (e.g., from the formulas of Table 4 or (8)-(10)); $\psi_{Fe}$ is the volume fraction of metallic iron in the oxide scale; $\lambda_{ox}$ is the total thermal conductivity coefficient of the oxide matrix (see (24)).

It is easy to see that the proposed chain of formulas "works" in the absence of metallic iron in the oxide scale. Then $\psi_{Fe} = 0$, and the successive application of formulas (21)-(25) gives the equality obvious in this case: $\lambda_{sc} = \lambda_{ox}$.

As an example, the graphs of $\lambda_{sc}$, which are calculated by the formulas (21)-(25) for the four hypothetical compositions of oxide scale listed in **Tables 5**, are shown in **Fig. 7**. In the first three cases, the same content of components in the entire temperature range (which in the first approximation can be regarded as corresponding to the conditions of rapid cooling) is conventionally assumed, and in the fourth - variable content depending on the temperature (simulating the conditions of slow cooling). In all cases, the oxides of the alloying elements are not taken into account. It can be seen that the main changes in the thermal conductivity of scale are associated with metallic iron. Thermal conductivity of scale consisting only of iron oxides (composition 1) varies in a narrow range from 3 to 6 W·m$^{-1}$K$^{-1}$. Increasing the proportion of metallic iron up to 12% leads to an increase in the thermal conductivity of scale up to 13-15 W·m$^{-1}$K$^{-1}$. This is especially evident for composition 4, when wüstite decays into magnetite and iron upon cooling.

**Table 5.** Oxide scale hypothetical compositions for the graphs in Fig. 7.

| Composition | Volume fraction of component $\psi$ | | | |
|---|---|---|---|---|
| | $Fe_{1-x}O$ | $Fe_3O_4$ | $Fe_2O_3$ | Fe |
| 1 | 0,8 | 0,15 | 0,05 | 0 |
| 2 | 0,5 | 0,35 | 0,1 | 0,05 |
| 3 | 0,2 | 0,55 | 0,15 | 0,1 |
| 4 | changes with temperature according to **Fig. 8** | | | |

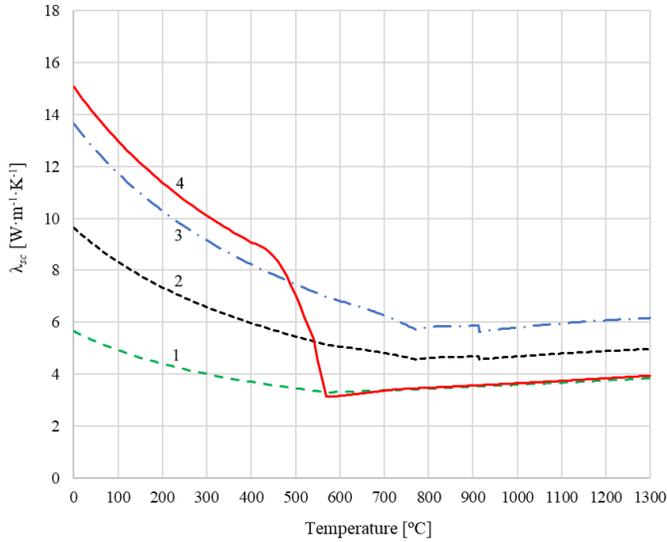

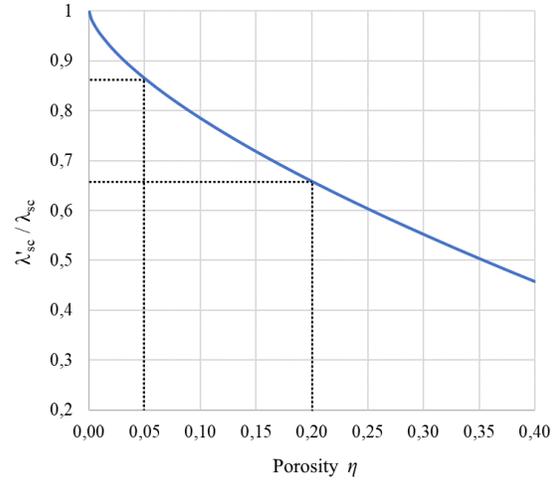

**Fig. 7.** Graphs of temperature dependence of thermal conductivity of scale calculated by formulas (21)-(25). Numbers in the curves are the numbers of calculated compositions of oxide scale according to Table 5.

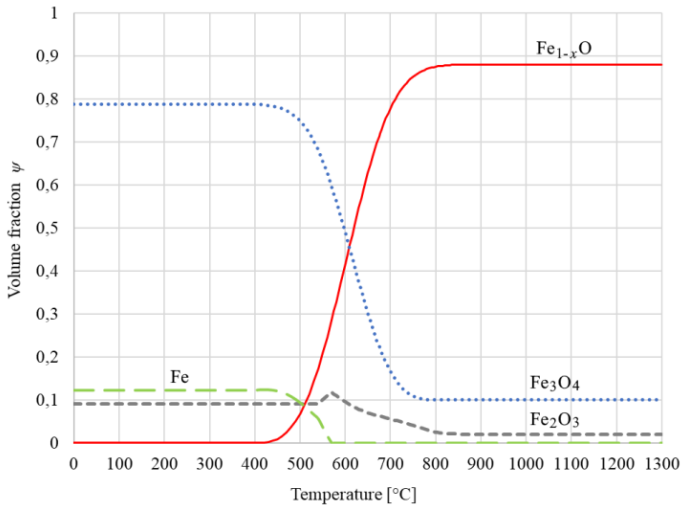

**Fig. 8.** Hypothetical variable oxide scale composition for simulations (number 4 in Fig. 7).

The influence of oxide scale porosity on its thermal conductivity coefficient can be considered according to a theoretically justified formula recommended in [27, p. 249] for porous materials with small, uniformly distributed air inclusions:

$$\lambda'_{sc} = \lambda_{sc}\left(1 - \eta^{2/3}\right) \qquad (26)$$

where $\lambda'_{sc}$ is the effective thermal conductivity coefficient of oxide scale with pores taken into account; $\lambda_{sc}$ is its true thermal conductivity coefficient without considering pores (for example, by (21)-(25)); $\eta$ is the porosity of oxide scale in fractions of one, (ratio of the pore volume to the total volume with pores).

**Fig. 9** shows the ratio of $\lambda'_{sc}/\lambda_{sc}$ as a function of oxide scale porosity, calculated from (26). The calculation for hot rolling conditions shows a decrease in thermal conductivity due to porosity for furnace scale ($\eta \geq 0.2$) by at least 35%, for air scale ($\eta \approx 0.05$) - by 15%.

**Fig. 9.** Relative coefficient of influence of oxide scale porosity $\eta$ on the value of its effective thermal conductivity coefficient (calculation by formula (26)). Dotted lines indicate data corresponding to the level of porosity of furnace and secondary scale during hot rolling.

## CONCLUSION

In order to obtain generalized formulas for calculating the thermal conductivity coefficient for iron oxides and pure iron, the empirical data of different authors are presented in the form of the inverse value - the specific heat resistance. As reference points for approximation we used two nodal points beyond the target range from 0 to 1300 ºC and the critical points connected with sharp change of properties: for magnetite $Fe_3O_4$ and hematite $Fe_2O_3$ - Curie points, for wüstite $Fe_{1-x}O$ - the boundary of its thermodynamic stability (Chaudron point), for pure iron Fe - Curie point and the temperature of polymorphic $\alpha \leftrightarrow \gamma$ transformation. As a result, generalized formulas were obtained that allow the values of critical temperatures and coordinates of the reference points to be varied. Taking into account the large scatter of empirical data, the accepted basic values of thermal conductivity at the reference points may be subject to correction in specific studies. In this case, the proposed approach makes it possible to automatically consider the new coordinate values in the estimates of the temperature dependence of the thermal conductivity coefficient while preserving the general form of the approximating functions.

The influence of different components of scale on its total thermal conductivity is proposed to consider in a combined way: metallic iron - by the Odelevski formula for chaotically located closed inclusions, and oxides - by the model of a sequential arrangement of layers perpendicular to the direction of the heat flow. Model calculations show that in the absence of metallic iron in oxide scale composition, its total thermal conductivity without considering pores can vary in a narrow range from 3 to 6 W·m$^{-1}$K$^{-1}$, depending on the temperature and the percentage of oxides. With an increase in fraction of metallic iron in the oxide scale up to 12% (e.g., as a result of eutectoid decomposition of wüstite), its thermal conductivity coefficient can increase to 15 W·m$^{-1}$K$^{-1}$.

According to the adopted assessment it is shown that with a pore fraction of 5% (which is typical for air scale) its effective thermal conductivity decreases by 15%, and with a pore fraction of 20% (which is typical for furnace scale during hot rolling) - by 35%.

The proposed methods are recommended for use in the mathematical modeling of the production and processing of steel products in the presence of surface oxide scale.